\documentclass[sigconf]{acmart}

\usepackage{subfigure}
\usepackage{multirow}
\usepackage{enumitem}
\usepackage{pifont}
\usepackage{xurl}

\usepackage{mathtools}
\usepackage{amsfonts}

\usepackage[linesnumbered,ruled,vlined]{algorithm2e}
\usepackage{graphicx}
\usepackage{textcomp}
\usepackage{xcolor}
\usepackage[prefixes-as-symbols=false]{siunitx}

\AtBeginDocument{%
  }

\setcopyright{acmcopyright}
\copyrightyear{2024}
\acmYear{2024}

\setcopyright{licensedusgovmixed}\acmConference[ISLPED '24]{ACM/IEEE International Symposium on Low Power Electronics and Design}{August 5--7, 2024}{Newport Beach, CA, USA}
\acmBooktitle{ACM/IEEE International Symposium on Low Power Electronics and Design (ISLPED '24), August 5--7, 2024, Newport Beach, CA, USA}
\acmDOI{10.xxxx/xxxxxxx.11111108}
\acmISBN{999-99-9999-9999-1/19/99}

\begin{document}

\title{OFHE: An Electro-Optical Accelerator for Discretized TFHE}

\author{
\centering
\setlength{\tabcolsep}{2pt}
\begin{tabular}{ccccccc}
Mengxin Zheng$^{\dag\S}$ & Cheng Chu$^\dag$ & Qian Lou$^\S$ & Nathan Youngblood$^*$ & Mo Li$^\ddag$ & Sajjad Moazeni$^\ddag$ &Lei Jiang$^\dag$
\end{tabular}
\setlength{\tabcolsep}{1pt}
\begin{tabular}{cccc}
\multicolumn{2}{c}{$^\dag$ Indiana University Bloomington} &\multicolumn{2}{c}{$^\S$ University of Central Florida} \\
\multicolumn{2}{c}{$^*$ University of Pittsburgh}          &\multicolumn{2}{c}{$^\ddag$ University of Washington} \\
\multicolumn{2}{c}{$^\dag$ \textit{\{zhengme, chu6, jiang60\}@iu.edu}}  &\multicolumn{2}{c}{$^\S$ \textit{qian.lou@ucf.edu}} \\
\multicolumn{2}{c}{$^*$ \textit{nathan.youngblood@pitt.edu}} &\multicolumn{2}{c}{$^\ddag$ \textit{\{moli96, smoazeni\}@uw.edu}} 
\end{tabular}
}

\renewcommand{\authors}{Mengxin Zheng, Cheng Chu, Qian Lou, Nathan Youngblood, Mo Li, Sajjad Moazeni, and Lei Jiang}
\renewcommand{\shortauthors}{M. Zheng et al.}

\begin{abstract}
This paper presents \textit{OFHE}, an electro-optical accelerator designed to process Discretized TFHE (DTFHE) operations, which encrypt multi-bit messages and support homomorphic multiplications, lookup table operations and full-domain functional bootstrappings. While DTFHE is more efficient and versatile than other fully homomorphic encryption schemes, it requires 32-, 64-, and 128-bit polynomial multiplications, which can be time-consuming. Existing TFHE accelerators are not easily upgradable to support DTFHE operations due to limited datapaths, a lack of datapath bit-width reconfigurability, and power inefficiencies when processing FFT and inverse FFT (IFFT) kernels. Compared to prior TFHE accelerators, OFHE addresses these challenges by improving the DTFHE operation latency by 8.7\%, the DTFHE operation throughput by $57\%$, and the DTFHE operation throughput per Watt by $94\%$.
\end{abstract}


\maketitle

\section{Introduction}
\label{s:intro}

The realm of cryptography has evolved significantly with the advent of \textit{Fully Homomorphic Encryption} (FHE)~\cite{Brakerski:TCT2014}. Tailored for cloud computing, FHE allows users, like Alice, to send encrypted data as ciphertexts to a server. The server can directly compute on these ciphertexts, ensuring unparalleled data privacy. Once processed, results, still encrypted, are returned to Alice. The server never accesses actual data, underscoring FHE's end-to-end encryption strength. Only Alice, using her secret key, deciphers the results.

\setlength{\abovecaptionskip}{2pt}
\setlength{\textfloatsep}{2pt}
\begin{table}[t!]
\centering
\caption{The comparison between various FHE schemes.}
\footnotesize
\setlength{\tabcolsep}{3pt}
\begin{tabular}{|c||c|c|c|c|}\hline
Scheme                                 & Operation Type         & Data Type     & Bootstrapping & Application \\\hline\hline
CKKS~\cite{Cheon:CEA2020}              & $\times$/$+$            & fixed-point  & high latency  & machine learning  \\\hline
TFHE~\cite{Chillotti:JC2018}           & logic                   & binary       & low latency  & general purpose     \\\hline\hline
DTFHE~\cite{Chillotti:ASIACRYPT2021}   & $\times$/$+$/logic/mod  & integer      & low latency  & general purpose \\\hline
\end{tabular}
\label{t:t_fhe_schemes}
\end{table}

In comparison to other FHE schemes, \textit{Discretized TFHE} (DTFHE) \cite{Chillotti:ASIACRYPT2021} emerges as a proficient FHE scheme, as illustrated in Table~\ref{t:t_fhe_schemes}.
\begin{itemize}[leftmargin=*, nosep, topsep=0pt, partopsep=0pt]
\item \textit{DTFHE exhibits versatility in operational types}. Traditional FHE schemes, e.g., CKKS~\cite{Cheon:CEA2020}, which only support homomorphic additions and multiplications, are potent in machine learning but lack applicability in general-purpose scenarios~\cite{Kotaro:USENIX2021}. While third-generation FHE schemes like TFHE~\cite{Chillotti:JC2018} focus on Boolean logic operations, they do not support native homomorphic additions and multiplications. For example, TFHE requires a lengthy sequence of logic gates to implement a multi-bit multiplication. In contrast, DTFHE enhances its capabilities to support homomorphic operations including Boolean algebra, modular arithmetic, and native multi-bit additions and multiplications. Compared to TFHE running on the same CPU, DTFHE reduces the latency of a 4-bit homomorphic multiplication by 99.5\%~\cite{Guimaraes:CEA2022}.
    
\item \textit{DTFHE can achieve fast bootstrapping operations}. In the context of FHE, operations performed on ciphertext inherently introduce noise. Over time, this accumulation of noise can impede accurate decryption. To mitigate this noise accumulation, FHE schemes invariably require periodic bootstrapping. Traditional FHE schemes like CKKS, unfortunately, suffer from prolonged bootstrapping durations, sometimes extending to several hundred seconds~\cite{Cheon:CEA2020}. However, akin to TFHE, DTFHE boasts rapid bootstrapping for binary messages. Nonetheless, for multi-bit integer messages, DTFHE's bootstrapping may require $\sim5s$ on a CPU. The expedited bootstrappings render DTFHE capable of implementing various general-purpose applications requiring substantial circuit depth.
\end{itemize}

Recent advancements in hardware accelerators~\cite{Jiang:DAC2022,Putra:MICRO2023} have been instrumental in expediting the bootstrapping operations of TFHE. Nevertheless, the domain of accelerating DTFHE—distinctive for its encryption of multi-bit messages within a singular ciphertext—remains largely unexplored. The challenge of adapting pre-existing TFHE accelerators to be compatible with DTFHE arises from two fundamental issues.
\begin{itemize}[leftmargin=*, nosep, topsep=0pt, partopsep=0pt]
\item The initial concern is the static nature of the datapaths in earlier TFHE accelerators, either 32~\cite{Jiang:DAC2022} or 64~\cite{Putra:MICRO2023} bits, which are ill-equipped to accommodate the versatile computational demands intrinsic to DTFHE. As an illustration, DTFHE's encryption of a singular-bit message can utilize a 32-bit datapath. However, for the homomorphic LUT operations on ciphertexts that encrypt messages up to 4 bits, a 64-bit datapath becomes imperative. Extending further, the homomorphic multiplications in DTFHE for ciphertexts with messages of 5 bits or more necessitate 128-bit polynomial multiplications, which subsequently calls for a 128-bit datapath. Regrettably, prior TFHE accelerators neither possess the infrastructure to sustain a 128-bit datapath nor the adaptability to alternate between diverse datapath bit-widths.

\item The second impediment revolves around the power efficiency—or lack thereof—in pre-existing TFHE accelerators. Their DSP~\cite{Beirendonck:ARIX2022} or CMOS ASIC~\cite{Jiang:DAC2022,Putra:MICRO2023} platforms for FFT and IFFT ((I)FFT\footnote{Henceforth, we will use (I)FFT to refer to both FFT and IFFT.}) are notorious energy consumers. Indeed, FFT and IFFT kernel processing is responsible for a $\sim55\%$~\cite{Jiang:DAC2022,Putra:MICRO2023} to $\sim80\%$~\cite{Nam:ICCAD2022} of the total energy expenditure in preceding TFHE accelerators. 
\end{itemize}

This paper unveils \textit{OFHE}, an electro-optical accelerator designed for DTFHE, with a focus on accelerating homomorphic multiplications and full-domain functional bootstrappings. The cornerstone of OFHE is a photonic FFT engine, engineered to speed up (I)FFT kernels across various scales and precisions integral to DTFHE operations, with high power efficiency. The other kernels integral to these operations are executed via CMOS modules. 
\begin{itemize}[leftmargin=*]
\item \textbf{A Chiplet Package Design}: At the heart of OFHE lies a photonic FFT engine adept at executing 64-point FFT functions. Leveraging passive photonic devices, the FFT engine's fabrication is optimized on dedicated chips. In parallel, CMOS modules find residence on a distinct CMOS chip. Each of these chips, whether photonic or CMOS, is organized as a individual chiplet, interconnected through photonic I/O links.

\item \textbf{Adaptable DTFHE Parameter Support}: OFHE's hallmark is its adaptability, accommodating a plethora of DTFHE parameters. Avoiding the slow runtime photonic device reconfiguration, OFHE incorporates a forward FFT engine, enriched with conjugating operations, tailored for IFFT kernels. Furthermore, a bit-level pipeline crafts a versatile 32-, 64-, or 128-bit (I)FFT datapath. A novel electro-optical computing flow segments a 1K-, 2K-, or 4K-point FFT kernel into several 64-point FFT kernels, which can be natively supported by the photonic FFT engine.

\item \textbf{Higher Throughput and Power Efficacy}: Experimental results show OFHE enhances the DTFHE operation latency by $8.7\%$, the DTFHE operation throughput by $57\%$ and the throughput per Watt by $94\%$, compared to prior TFHE accelerators.
\end{itemize}

\vspace{-0.1in}
\section{Background}
\label{s:back}

\subsection{Discretized TFHE (DTFHE) Basics}
\textbf{Notations}. In this paper, (1) boldface signifies vectors; (2) superscripts express the element count in vectors; (3) modulus is indicated by subscripts. (4) $\mathbb{S}_q^n$ denotes the set of $n$-element vectors in $\mathbb{S}$ modulo $q$. (5) $\mathbb{S}[X]$ represents the set of polynomials over variable $X$ with $\mathbb{S}$ coefficients. (6) Power-of-2 cyclotomic polynomial's modulo is portrayed by its degree. (7) Vectors of polynomials over $X$ with modulus $\Phi{2N}(X)=X^N+1$ and coefficients in $\mathbb{S}$ modulo $q$ are represented by $\mathbb{S}_q[X]_N^n$. (8) $\mathbb{M}$ stands for the $\mathfrak{R}$-module.


\textbf{Binary-Secret Scale-Invariant LWE}. FHE foundations lie in the Learning With Errors (LWE) problem. An LWE sample encompasses a duo $(\mathbf{a},b) \in \mathbb{M}^{n+1}$, with $\mathbf{a}$ uniformly drawn from $\mathbb{M}^n$, $b=\langle \mathbf{a}, \mathbf{s} \rangle+e \in \mathbb{M}$, and $n\geq1\in \mathbb{Z}$. The binary secret key $\mathbf{s}$ is uniformly selected from $\mathfrak{B}^n$. Meanwhile, error $e$ arises from a Gaussian distribution over $\mathbb{M}$, centralized at 0 with standard deviation $\sigma$.

\textbf{Encryption and Decryption}. The idea of a LWE-based cryptosystem is to encrypt a message by adding the message to the $b$ part of an LWE sample, since it is indistinguishable from a vector sampled from the uniform distribution. TFHE encrypts 1-bit messages in both TLWE and TRLWE samples~\cite{Chillotti:ASIACRYPT2021}. Both are a type of LWE samples, differing by the definition of $\mathbb{M}$ and $\mathfrak{B}$.

\SetKwInput{KwInput}{Input}                
\SetKwInput{KwOutput}{Output}              
\begin{algorithm}[t!]
\DontPrintSemicolon
\scriptsize
\KwInput{a TLWE sample $c=(\mathbf{a},b)\in \text{TLWE}_s(\frac{m}{B})$, $m\in \mathbb{Z}_B$; a LUT $\mathbf{L} = [l_0,\ldots,l_{B-1}]\in\mathbb{Z}_B^B$; a bootstrap. key $BK_i\in \text{TRGSW}_S(s_i)$, $i\in\llbracket1,n \rrbracket$}
\KwOutput{$c'\in\text{TLWE}_S(\frac{\mathbf{L}[m]}{B})$, $S\in \mathbb{B}^N$}
\SetKwFunction{FMain}{FullDomainBootstrap}
\SetKwProg{Fn}{Function}{:}{}
\Fn{\FMain{$(\mathbf{a},b)$, $\mathbf{L}$, $\mathbf{BK}$}}{\label{lst:line:FDFB}

$tv \leftarrow \sum_{i=0}^{\frac{B}{2}-1}\sum_{j=0}^{1}\sum_{k=0}^{\frac{N}{B}-1}\frac{1}{B}l_{\frac{jB}{2}+i} X^{(2i+j)\frac{N}{B}+k}$; $pa \leftarrow \frac{B+1}{4B}$ 

$c_{sign} \leftarrow$ FunctionalBootstrap($c$,$[pa,\ldots, pa]$, $\mathbf{BK}$) $-pa$

\Return FunctionalBootstrap($c+c_{sign}$, $tv$, $\mathbf{BK}$)
}

\SetKwFunction{FMain}{FunctionalBootstrap}
\SetKwProg{Fn}{Function}{:}{}
\Fn{\FMain{$(\mathbf{a},b)$, $\mathbf{L}$, $\mathbf{BK}$}}{\label{lst:line:functionalboot}

$b\leftarrow \lceil 2Nb \rfloor$ and $a_i\leftarrow 2Na_i\in \mathbb{Z}_{2N}$, $i\in\llbracket1,n \rrbracket$ 

$tv \leftarrow \sum_{i=0}^{N-1} \frac{1}{2B}\cdot l_{\lfloor \frac{iB}{N} \rfloor} X^i \in \mathbb{T}_N[X]$ \label{lst:line:functionencode}

$ACC \leftarrow \text{BlindRotate}((0,tv), (\mathbf{a},b+\frac{1}{4B}), \mathbf{BK}))$

\Return SampleExtract$(ACC)$
}

\SetKwFunction{FMain}{BlindRotate} 
\SetKwProg{Fn}{Function}{:}{}
\Fn{\FMain{$(\mathbf{a},b)$, $tv$, $\mathbf{BK}$}}{\label{lst:line:blindroate}

$ACC\leftarrow X^{-\lceil b2N \rfloor}\cdot tv$

\For{$i=1$ to $n$}
{
$ACC \leftarrow BK_i \cdot (ACC - X^{\lceil a_i2N \rfloor}\cdot ACC) + ACC$
}
\Return $ACC$

}
\caption{Various TFHE Bootstrappings.}
\label{a:tlight_tfhe_boot}
\end{algorithm}

\textbf{DTFHE}. DTFHE~\cite{Chillotti:ASIACRYPT2021} is proposed to encrypt a multi-bit message in a T(R)LWE sample. $c=(\mathbf{a}, b)\in \text{TLWE}(\frac{m}{2B})$ is a TLWE sample, where $m$ indicates a multi-bit message, $B$ denotes a discretization parameter, and only a half of the torus is used. DTFHE encodes the message by $m=\sum_{i=0}^{\lfloor \log_2 m\rfloor} 2^i \tilde{m}_i$, where $\tilde{\mathbf{m}}$ is the binary vector representation of $m$. Messages are mapped to integers by discretizing the torus. $B$ specifies the base in which messages are decomposed when working with messages encrypted in multiple samples. 

\vspace{-0.1in}
\subsection{DTFHE Operations and Implementations}

\textbf{Gate Bootstrapping}. At the end of a two single-bit inputs TFHE gate, a gate bootstrapping is required to remove the noises accumulated in the ciphertext. A gate bootstrapping can be summarized in three steps: (1) setting the accumulator vector, $ACC$, to be $\sum_{i=0}^N \frac{1}{4} X^i \in \mathbb{T}_N[X]$; (2) using \textit{BlindRotate} to compute $ACC\cdot X^{-\phi(c)2N}$ mod $\Phi_{2N}$; and (3) using \textit{SampleExtract} to extract the constant term of the rotated $ACC$. Particularly, the most important step, \textit{BlindRotate}, in a gate bootstrapping is shown in Line~\ref{lst:line:blindroate} of Algorithm~\ref{a:tlight_tfhe_boot}, where a blind rotation of $ACC$ by $-\phi(c)$ is performed. 

\textbf{Functional Bootstrapping}. Besides removing noises, a functional bootstrapping~\cite{Putra:MICRO2023} also performs a homomorphic lookup table (LUT) through replacing the regular test vector by an encoded LUT, as shown in Line~\ref{lst:line:functionalboot} of Algorithm~\ref{a:tlight_tfhe_boot}. A functional bootstrapping discretizes the domain of a function, evaluates the function in all discretized points, and then stores the results in a LUT. The LUT is encoded as a polynomial (Line~\ref{lst:line:functionencode}, where $B$ is a discretization parameter). Due to the negacyclic property of \textit{BlindRotate}, the function supported by functional bootstrapping has to be anti-symmetric, i.e., $f(x+N)=-f(x)$. For an arbitrary function, the negacyclic property can be avoided using only the first half of the torus.

\textbf{Full-Domain Functional Bootstrapping (FDFB)}. Besides using only the first half of the torus, a functional bootstrapping cannot perform homomorphic modular arithmetic. FDFB~\cite{Guimaraes:CEA2022} shown in Line~\ref{lst:line:FDFB} of Algorithm~\ref{a:tlight_tfhe_boot} is created to work over the entire torus and to support homomorphic modular arithmetic. A FDFB combines two functional bootstrappings in a chaining way, which guarantees the lowest output error variance.

\textbf{TRLWE Tensor Product and TLWE multiplication}. DTFHE supports BFV-like TRLWE tensor product operations~\cite{Chillotti:ASIACRYPT2021} by native TFHE parameters. A integer parameter $q$ controls the precision of the TRLWE result. A TRLWE tensor product also requires a relinearization key to reduce the number of terms in the TRLWE result to 2. Via a TRLWE tensor product, DTFHE can support a homomorphic multiplication between two TLWE samples. When implementing an $m$-bit $\times$ $n$-bit integer homomorphic multiplication, a DTFHE TRLWE tensor product is faster than $\mathcal{O}(mn)$ TFHE gates.

\textbf{Torus Implementation}. In order to map and rescale all elements from $\mathbb{T}$ to $\mathbb{Z}_q$, an integer precision parameter $q\in\mathbb{N}$ is used when implementing a torus. For TFHE, which encrypts 1-bit messages and works with two single-bit inputs logic gates, $q=2^{32}$~\cite{Chillotti:ASIACRYPT2021} is precise enough. However, DTFHE requires a larger plaintext space to encrypt multiple bits, perform homomorphic multiplications, and support full-domain functional bootstrappings. So DTFHE uses $q=2^{64}$~\cite{Chillotti:ASIACRYPT2021} to map all torus elements to 64-bit long integers.

\begin{figure}[t!]
\centering
\begin{minipage}{.22\textwidth}
\centering
\includegraphics[width=1.53in]{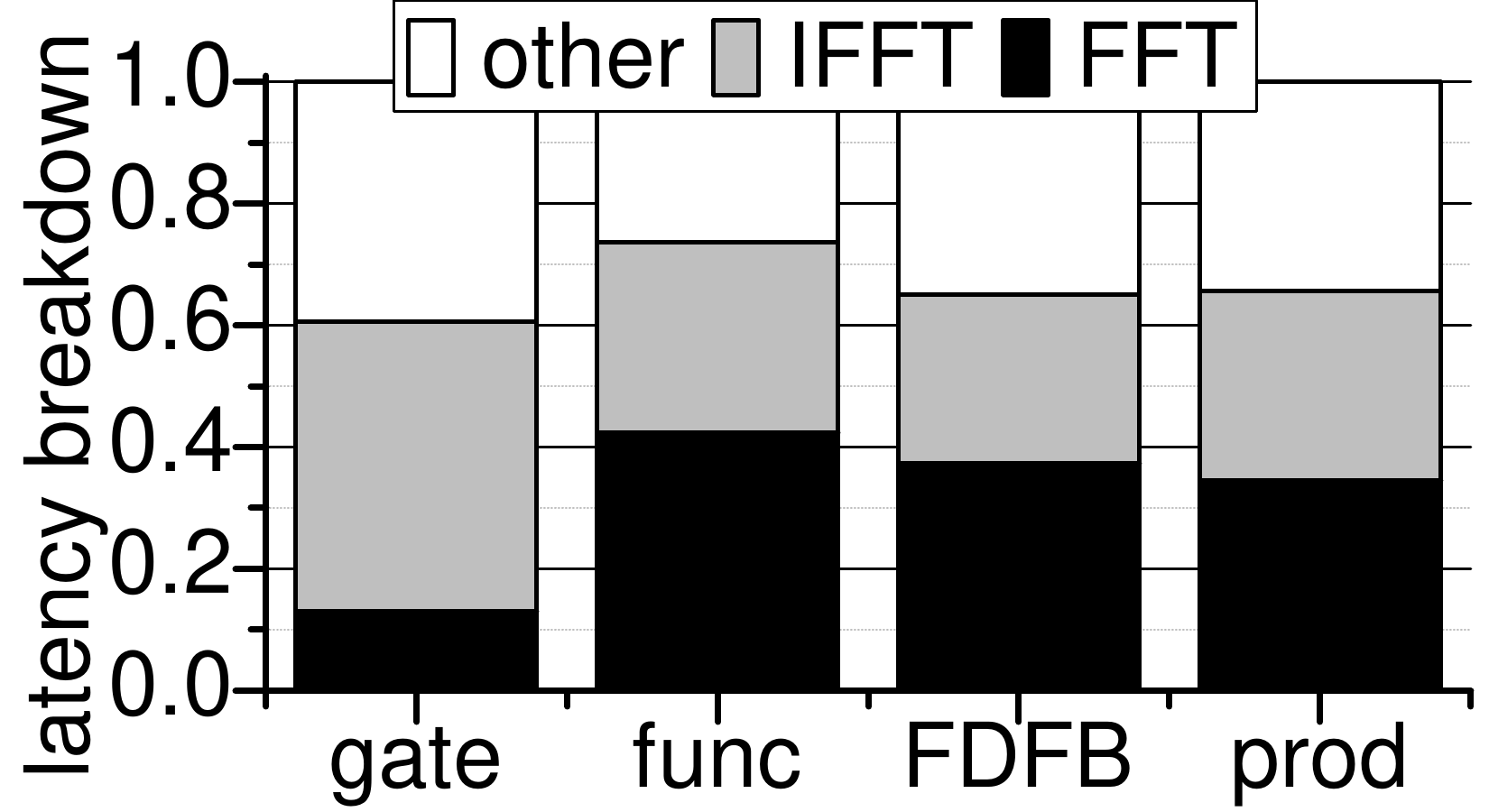}
\caption{The latency breakdown of DTFHE operations (i.e., gate, functional, full-domain bootstrappings, and TRLWE tensor product).}
\label{f:tlight_op_breakdown}
\end{minipage}
\hspace{0.1in}
\begin{minipage}{.22\textwidth}
\centering
\includegraphics[width=1.53in]{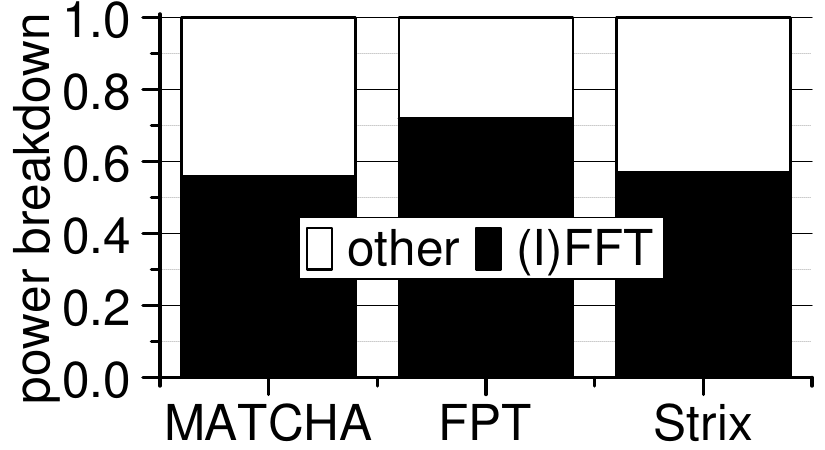}
\caption{The power breakdown of prior TFHE hardware accelerators including MATCHA~\cite{Jiang:DAC2022}, FPT~\cite{Beirendonck:ARIX2022} and Strix~\cite{Putra:MICRO2023}.}
\label{f:tlight_op_hardware}
\end{minipage}
\end{figure}

\textbf{(I)FFT}. Homomorphic multiplications and various types of bootstrappings in DTFHE require numerous polynomial multiplications~\cite{Chillotti:ASIACRYPT2021}. To reduce the computational complexity of polynomial multiplication from $\mathcal{O}(N^2)$ to $\mathcal{O}(N \log N)$, where $N$ is the degree of the polynomials, DTFHE uses (I)FFT. As shown in Figure~\ref{f:tlight_op_breakdown}, (I)FFT kernels consume $64\%\sim75\%$~\cite{Jiang:DAC2022} of the latency of various DTFHE operations running on a CPU. Our experimental methodology is explained in Section\ref{s:expmeth}. When the integer precision parameter $q$ is $2^{32}$, 64-bit double-precision floating-point (I)FFT kernels can be directly used without introducing significant errors. However, if $q=2^{64}$, particularly when the numeric base $B$ in Algorithm~\ref{a:tlight_tfhe_boot} is large, polynomial multiplications of DTFHE require up to 128-bit precision. Therefore, double-precision floating-point (I)FFT kernels are not precise enough for various DTFHE operations~\cite{Guimaraes:CEA2022}.

\vspace{-0.2in}
\subsection{Related Work and Motivation}
\label{s:related}

\textbf{Related Work}. FHE is renowned for its security but often criticized for its high computational overheads. To address these challenges, ASIC-based accelerators~\cite{Kim:ISCA2022,Zheng:NANOARCH2022} have been developed for FHE schemes such as CKKS and BGV, which support only homomorphic multiplications and additions. Additionally, several accelerators~\cite{Jiang:DAC2022,Putra:MICRO2023} have emerged, focusing on bootstrapping functions within single-bit TFHE ciphertexts. However, no accelerator adequately supports DTFHE encrypting multi-bit messages. Although prior work~\cite{Zheng:NANOARCH2022} leverages photonic microdisks to design 512-bit electro-optical adders and multipliers, enhancing BGV's number-theoretic transform (NTT) processes, no existing work except OFHE uses analog photonic signals to accelerate (I)FFT kernels for DTFHE.

\textbf{Motivation}. Existing TFHE accelerators have limitations when processing DTFHE operations. Two constraints are observed:
\begin{itemize}[leftmargin=*, nosep, topsep=0pt, partopsep=0pt]

\item \textbf{Limited Datapath Scalability}: Efforts to retrofit current TFHE accelerators~\cite{Jiang:DAC2022,Putra:MICRO2023} for DTFHE encounter several barriers. Intrinsic operations of DTFHE, like full-domain functional bootstrappings, TRLWE tensor products, and TLWE multiplications, necessitate 128-bit polynomial computations. Most prior accelerators offer datapaths confined to 32 bits, thus remaining incompatible with highly precise (I)FFT kernels or 128-bit polynomial computations. While DTFHE's single-bit message encryption parallels TFHE, demanding a 32-bit datapath, multi-bit (e.g., 4-bit) encryption in DTFHE necessitates a 64-bit one. To adeptly manage these variable requirements, accelerators must be equipped with flexible data paths of 32, 64, and 128 bits, a feature conspicuously absent in current TFHE accelerators.

\item \textbf{Power Efficiency Concerns}. Predominantly prior TFHE accelerators~\cite{Jiang:DAC2022,Beirendonck:ARIX2022,Putra:MICRO2023} are notorious power guzzlers, with consumption often breaching $40\sim100$ Watts. In these accelerators, (I)FFT kernels notably surge the power demands, accounting for nearly 56\% to 80\% of their total consumption, as depicted in Figure~\ref{f:tlight_op_hardware}.
\end{itemize}

\vspace{-0.1in}
\section{OFHE}
\label{s:tlight}

\begin{figure*}[t!]
\centering
\includegraphics[width=6.6in]{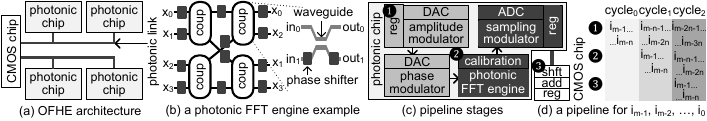}
\caption{The components and pipeline of OFHE, when processing $m$-bit inputs $i_{m-1},i_{m-2},\ldots,i_{0}$.}
\label{f:tlight_optical_all222}
\vspace{-0.2in}
\end{figure*}

\textbf{Architecture}. This paper presents OFHE, a photonic accelerator designed for efficient DTFHE operations. OFHE comprises multiple photonic FFT chips and a CMOS chip, as depicted in Figure~\ref{f:tlight_optical_all222}(a). The components are structured as chiplets interconnected through photonic I/O links. The CMOS chip initializes and calibrates all photonic components. When an FFT or IFFT kernel is invoked, the CMOS chip transmits input data to a photonic chip using an I/O link. Digital-to-Analog Converters (DACs) transform this data into optical signals, which are processed by the photonic chip. The resultant signals are then sampled and digitized into final outputs through Analog-to-Digital Converters (ADCs).

\textbf{Photonic FFT}. Unitary FFT operations can be performed using optical devices~\cite{Hillerkuss:OE2010}: $\mathbf{X}_k=\frac{1}{\sqrt{N}}\sum_{n=0}^{N-1}\mathbf{x}_n e^{-i\frac{2\pi k n}{N}}$, where $k\in \llbracket0,N-1 \rrbracket$, $\mathbf{x}_n$ represents an input sampling point, and $\mathbf{X}_k$ denotes an output point. A 4-point Photonic FFT Engine (PFFTE) is shown in Figure~\ref{f:tlight_optical_all222}(b), where a butterfly unit is a 2-point unitary FFT substrate. The butterfly of the 2-point FFT  $\begin{pmatrix} out_0\\ out_1 \end{pmatrix} =\frac{1}{\sqrt{2}}\begin{pmatrix} in_0+in_1\\ in_0-in_1 \end{pmatrix}$ can be described by $\underbrace{\begin{pmatrix} 1 & 0 \\ 0 &-j \end{pmatrix}}_\text{output phase shifter} \underbrace{\frac{1}{\sqrt{2}}\begin{pmatrix} 1 &j\\ j &1 \end{pmatrix}}_\text{coupler} \underbrace{\begin{pmatrix} 1 & 0 \\ 0 &-j \end{pmatrix}}_\text{input phase shifter} \begin{pmatrix} in_0\\ in_1 \end{pmatrix}$, where $in_0$ and $in_1$ are inputs, and $out_0$ and $out_1$ are outputs. The first and third matrices are implemented by input and output phase shifters, respectively, while the second matrix is performed by a $2\times2$ directional coupler. All components of the 2-point PFFTE are passive optical devices, and therefore consume little power.


\textbf{Photonic Chip}. On a photonic chip, we use ITO plasmonic MZ modulators~\cite{Amin:IPRSN2020} as amplitude and phase modulators, and Michelson modulators as sampling modulators. To construct a PFFTE, various compact and low-loss photonic components are employed, including $2\times2$ directional couplers~\cite{Ye:NAN2015}, phase shifters~\cite{Nicholas:OE2014}, Y-splitters and combiners, and straight and spiral~\cite{Hong:PR2022} waveguides. The input/output interfaces of the PFFTE are implemented using grating couplers. The length of the spiral waveguide of a phase shifter in the $i_{th}$ FFT stage is calculated as $T/2^i$, where $1\leq i\leq \log_2(N)$, $N$ is the number of FFT inputs, and $T$ denotes the physical length of time delay. $T$ can be computed as $c/(n_{eff}*freq)$, where $c$ is the speed of light, $n_{eff}$ is the effective index, and $freq$ is the operating frequency of the PFFTE. We set $freq$ to 12GHz~\cite{Hillerkuss:OE2010}, resulting in $T=10$ mm.

\begin{figure}[t!]
\centering
\begin{minipage}[c]{0.22\textwidth}
\centering
\includegraphics[width=1.55in]{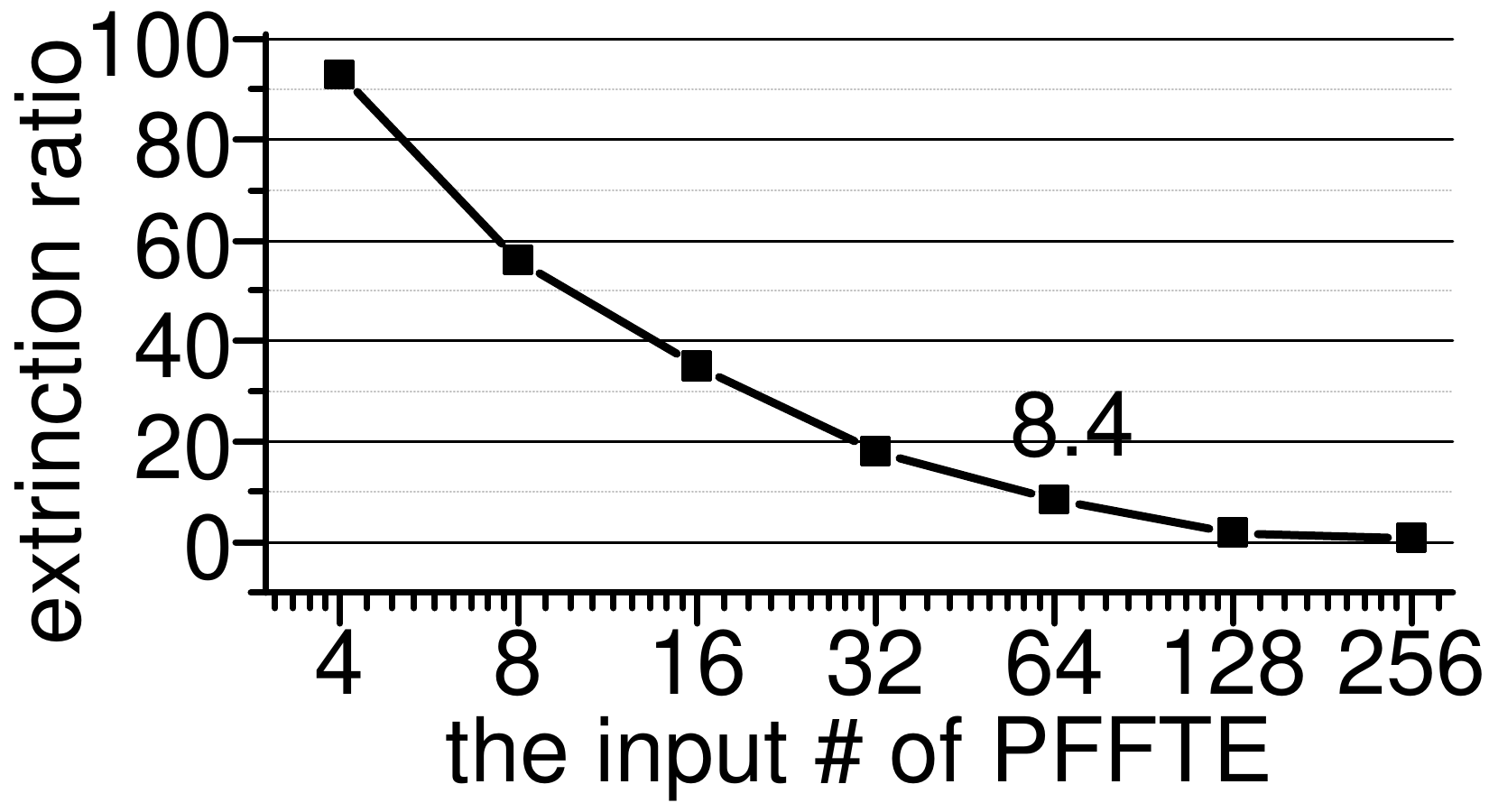}
\caption{PFFTE extinction ratio with varying input \#.}
\label{f:tlight_op_er}
 \end{minipage}
\hspace{0.1in}
\begin{minipage}[c]{0.22\textwidth}
 \centering
\includegraphics[width=1.55in]{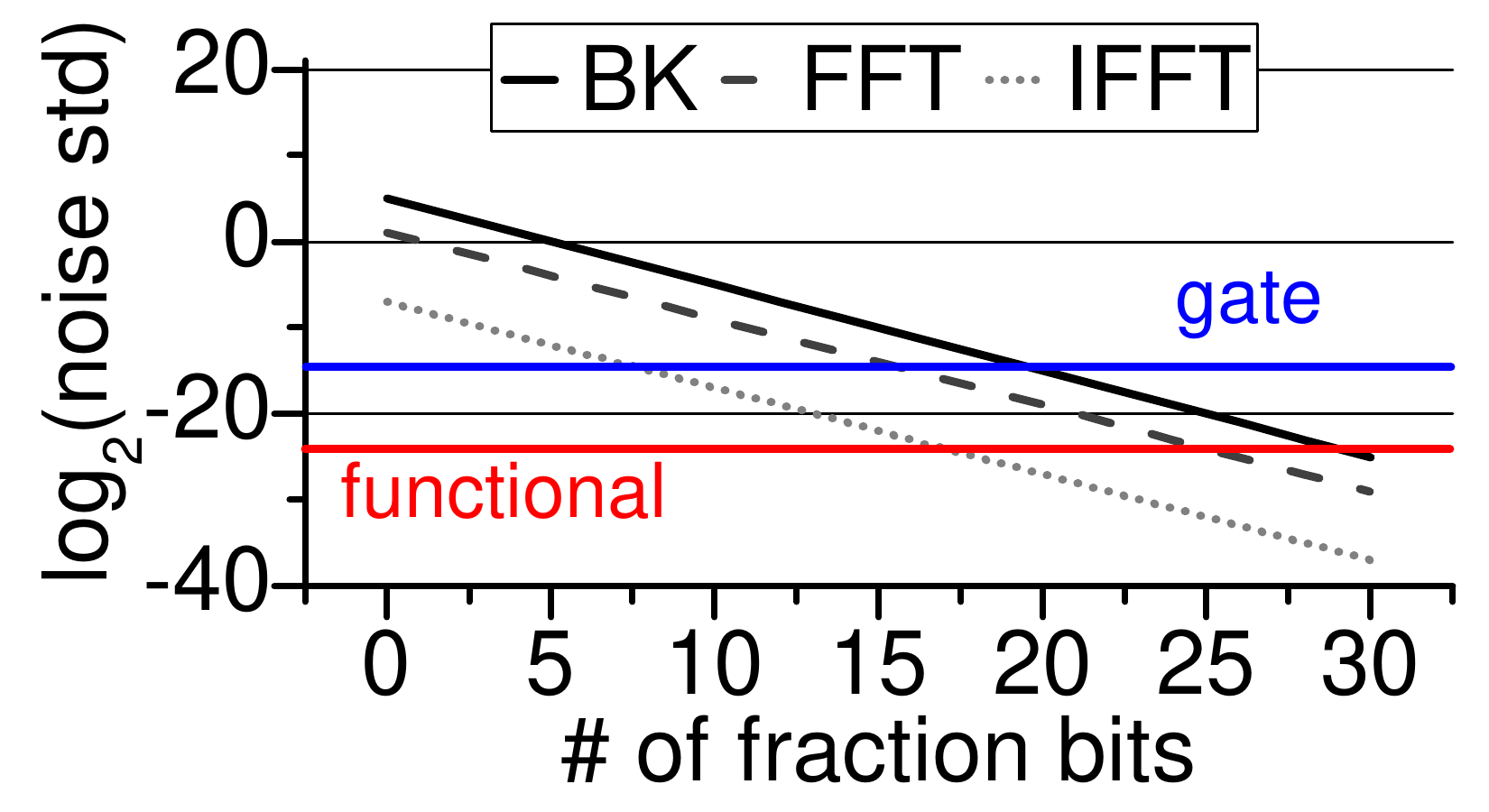}
\caption{Fraction bit \# vs output approx. noise.}
\label{f:tlight_fixed_point}
\end{minipage}
\end{figure}

\textbf{The Input Scale of a PFFTE}. As the integration of additional photonic components into a PFFTE escalates, there is an inevitable rise in total optical loss. This uptick demands compensation via heightened input power~\cite{Hillerkuss:OE2010}. Consequently, to avert disproportionate power usage, moderating the input count ($N$) becomes paramount. High values of $N$ accentuate the optical losses associated with spiral waveguides~\cite{Hillerkuss:OE2010}, making them the principal contributors to the PFFTE's optical loss. The extinction ratio of a PFFTE, expressed as the ratio $P_{max}/P_{min}$ (with $P_{max}$ and $P_{min}$ being the peak and nadir of PFFTE output powers respectively), diminishes significantly with ascending input numbers, as depicted in Figure~\ref{f:tlight_op_er}, due to the increased necessity for spiral waveguides. For precise sampling by modulators, maintaining a robust PFFTE output power is vital, hence a restricted input count for the PFFTE is advocated. Our design focuses on a 64-point PFFTE.

\textbf{Fixed Point Representation}. In TFHE and DTFHE, elements from $\mathbb{T}$ are mapped to $\mathbb{Z}_q$, with $q$ taking values $2^{32}$ or $2^{64}$. Previous TFHE accelerators, when $q=2^{32}$, leveraged 38-~\cite{Jiang:DAC2022} and 30-bit fixed-point results within (I)FFT cores, deviating from the conventional double floating-point computations. The 30-bit depiction as per Figure~\ref{f:tlight_fixed_point} divides into 19-, 14-, and 6-bit fractions for respective bootstrapping-key, FFT, and IFFT computations, thereby ensuring consistent gate bootstrapping activities. Contrarily, earlier DTFHE CPU executions employed the Karatsuba algorithm~\cite{Guimaraes:CEA2022} since achieving quadruple-precision floating-point (I)FFTs on CPUs posed challenges. OFHE, aiming for minimized hardware expenditure, translates the intermediary outcomes and twiddle factors of (I)FFT cores in different DTFHE operations into fixed-point representations. Adopting this scheme, OFHE cautiously utilizes 64-bit and 128-bit fixed-point representations for $q=2^{32}$ and $q=2^{64}$, respectively, dividing them into 29-, 26-, and 17-bit fractions.

\textbf{FFT Pipeline with a Reconfigurable Datapath}. DACs and ADCs are limited by their precision, making it infeasible for a PFFTE to process integer inputs of $m$-bit in a singular operation, with $m$ being $32$, $64$, or $128$ bits. Thus, OFHE segments the FFT operation into a tri-stage pipeline, as depicted in Figure~\ref{f:tlight_optical_all222}(c). In stage \ding{182}, DACs extract $n_{in}$ bits from the FFT input register, converting these to photonic signals transmitted to the PFFTE. Here, $n_{in}$ specifies the DAC's bit-width. In stage \ding{183}, the PFFTE's output signals are relayed to sampling modulators. ADCs then transform these voltage outputs into $n_{out}$-bit digital FFT outcomes, subsequently stored in a register, where $n_{out}$ denotes the ADC's bit-width. In stage \ding{184}, a shifting and addition procedure on the CMOS chip compiles each $n_{out}$-bit FFT outcome into an $m$-bit conclusive result. An example pipeline, initialized from input's most significant bits, is portrayed in Figure~\ref{f:tlight_optical_all222}(d). Achieving a stable pipeline kernel necessitates three cycles from the PFFTE. The FFT pipeline's bit-width adaptability is facilitated by varying $m$ values, with its operational frequency tethered to the ADCs' and DACs' speeds.

\textbf{Fast IFFT}. Although it is possible for the PFFTE to perform an IFFT operation through reconfiguring phase shifters, this approach has a long latency and requires a large number of DACs. Reconfiguring a phase shifter can take several milliseconds~\cite{Hillerkuss:OE2010}, and adjusting the parameters of all necessary phase shifters concurrently for an IFFT operation requires hundreds of DACs. To avoid the expensive run-time phase shifter reconfiguration, OFHE employs the conversion~\cite{Duhamel:TASSP1988} between IFFT and FFT. The conversion is described as
$\text{IFFT}(\mathbf{X}_k)=\frac{1}{N}\text{conj}(\text{FFT}(\text{conj}(\mathbf{X}_k)))$, where $k\in \llbracket0,N-1 \rrbracket$, and $\text{conj}$ represents a conjugate operation. In this way, OFHE can perform conjugate operations on the input and output points of an IFFT operation on the CMOS chip and use the same PFFTEs to complete the IFFT operation.



\textbf{Scaling FFT}. To work with different DTFHE parameters with assorted security levels, OFHE needs to execute FFT operations on inputs spanning 1K, 2K, and 4K points. Given that PFFTE inherently supports only 64-point FFT operations, OFHE employs the 4-step FFT algorithm~\cite{Bailey:SUPER1989}. This algorithm compartmentalizes extensive FFT operations into numerous 64-point FFT tasks. Specifically, for an $N$-point FFT (where $N$ encompasses 1K, 2K, or 4K), OFHE conducts $n_0$ 64-point FFT operations on the input points structured as an $n_0\times64$ matrix, leading to a resultant matrix $R_{pq}$. Subsequently, this matrix undergoes multiplication by the exponential factor $e^{-2\pi ipq/N}$ and is transposed. Ultimately, OFHE executes 64 $n_0$-point FFT operations on the resulting matrix via the PFFTE, with necessary padding introduced when $n_0$ is less than 64.


\textbf{CMOS Chip}. The CMOS chip, interfacing with PFFTE chips via photonic I/O links, comprises 8 cores that manage computations for DTFHE operations, excluding (I)FFT kernels. A crossbar network connects these cores with a scratchpad memory. One core uses serial adders to collate intermediate results derived from the PFFTEs. Each core is calibrated to support datapaths of 32, 64, and 128 bits, leveraging serial adders and multipliers for polynomial additions and element-wise multiplications. Conjugating units modify IFFT inputs for direct PFFTE processing. Furthermore, a transposition and shuffling mechanism optimizes the data structure for inputs and intermediate (I)FFT kernel results. 1KB input and output registers serve all computational units within a core.

\begin{table}[t!]
\centering
\caption{The area and power consumption of OFHE}
\setlength{\tabcolsep}{3pt}
\footnotesize
\begin{tabular}{|c||l|l|l|l|}\hline
Name                          & Component          & Spec                                  & Power ($mW$)  & Area ($mm^2$) \\\hline\hline

                              & D. coupler         & \cite{Ye:NAN2015}, $\times 192$       & 0             & 0.0013 \\\cline{2-5}
          									  & P. shifter         & \cite{Nicholas:OE2014}, $\times 448$  & 0   	         & 0.014  \\\cline{2-5}
                              & spiral             & \cite{Hong:PR2022}, $\times 192$      & 0             & 0.06  \\\cline{2-5}
           										& splt/comb          & $\times 484$                          & 0             & 0.028  \\\cline{2-5}
          					          & G. coupler         & $\times 128$                          & 0             & 0.003  \\\cline{2-5}
64-point  						        & MZ mod.            & \cite{Amin:IPRSN2020}, $\times 64$    & 0.121         & 0.0069  \\\cline{2-5}
photonic											& samp. mod.         & $\times 64$                           & 4.096         & 0.0076  \\\cline{2-5}
FFT chip                      & holder             & 12G, $\times 64$                      & 0.001         & 0.0001  \\\cline{2-5}
															& 1-bit DAC          & 12G, $\times 128$                     & 2.5           & 0.0001 \\\cline{2-5}
                              & 6-bit ADC          & \cite{adc_survey}, 12G, $\times 64$   & 1280          & 0.768 \\\cline{2-5}
															& I/O buff.          & 32KB, $\times1$                       & 9.92    	     & 0.017  \\\cline{2-5}
                							& laser              & $\times1$                             & 130           & 0.235 \\\cline{2-5}
                							& chplt PHY          & 1KB                                   & 19.2		       & 0.864   \\\cline{1-5}
sub-total                     &                    &                                       & 1.45 Watt     & 1.99   \\\hline\hline
                 							& I/O buff.          & 4KB, $\times8$                        & 9.92          & 0.017 \\\cline{2-5}
32nm                          & core               & 1.2G, $\times8$                       & 15,920        & 13.76 \\\cline{2-5}
CMOS                          & NoC                & $8\times32$, 256b                     & 2110          & 0.44 \\\cline{2-5}
chip                          & SPM                & 2MB, 32 banks                         & 1510          & 1.64 \\\cline{2-5}
															& chplt PHY          & 1KB                                   & 19.2          & 1.728 \\\cline{2-5}
															& mem ctrl           & DDR5 PHY                              & 6,400	       & 7.5  \\\cline{1-5}
sub-total                     &                    &                                       & 26.03 Watt    & 30.66   \\\hline\hline
\multirow{2}{*}{OFHE}         & OFFT chip          & $\times8$                             & 11.6 Watt     & 15.92 \\\cline{2-5}
															& CMOS chip          & $\times2$                             & 52.06 Watt    & 61.32 \\\cline{1-5}							
\textbf{total}                &                    &                                       & \textbf{63.66} Watt    & \textbf{77.2}   \\\hline									
\end{tabular}
\label{t:t_power_area}
\end{table}

\textbf{Design Overhead}. Table~\ref{t:t_power_area} presents the power and area overhead of OFHE. The system comprises 8 PFFTE chips, operating at 12 GHz, and two CMOS chips running at 1.2 GHz. 
\begin{itemize}[leftmargin=*, nosep, topsep=0pt, partopsep=0pt]
\item \textbf{PFFTE chip}. Each PFFTE chip is constructed using $\sim1.5K$ passive photonic components. The light for a PFFTE is generated by an on-chip laser. The PFFTE guarantees only 6-bit accurate FFT intermediate results, due to the relative intensity noise of our laser. These components are modeled using Lumerical FDTD and INTERCONNECT software tools. 

\item \textbf{CMOS chip}. Each CMOS chip is modeled using Cadence Virtuoso with 32nm PTM technology. Memory parts are modeled using CACTI. Each core has 1K serial adders, 1K serial multipliers, 1K conjugate units, four transpose and shuffling units. 
\end{itemize}


\vspace{-0.1in}
\section{Experimental Methodology}
\label{s:expmeth}

\textbf{Schemes}. OFHE is compared against four prior accelerators as shown in Table \ref{t:t_hard_comp}. The comparison includes FPT, an FPGA-based accelerator~\cite{Beirendonck:ARIX2022}, as well as two ASIC-based accelerators, MATCHA (MATC)~\cite{Jiang:DAC2022} and Strix~\cite{Putra:MICRO2023}. Additionally, CryptoLight (Cryp)~\cite{Zheng:NANOARCH2022}, an electro-optical accelerator originally designed for processing BGV by photonic NTT units, was selected. Since its detailed configurations are not reported in~\cite{Zheng:NANOARCH2022}, we adapted Cryp to accelerate DTFHE by replacing the CMOS adders and multipliers of Strix with the electro-optical counterparts of Cryp while following the same configuration of Strix. It is noted that all prior TFHE accelerators support only gate-level functional bootstrappings (GFBs) and are unable to perform BFV-like TLWE multiplications or full-domain functional bootstrappings (FDFBs).

\textbf{DTFHE Operations and Parameters}. To compare OFHE against prior TFHE accelerators, we studied GFBs operating on 32-bit datapaths. We also studied DTFHE-specific FHE operations including BFV-like TLWE multiplications, and FDFBs, which may require 64- or 128-bit datapaths. As Table~\ref{t:t_tfhe_para} shows, we studied 4 sets of DTFHE parameters, i.e., I and II typically for GFBs, while III and IV~\cite{Guimaraes:CEA2022} often for TLWE multiplications and FDFBs.

\begin{table}[t!]
\centering
\caption{The comparison of TFHE hardware accelerators.}
\footnotesize
\setlength{\tabcolsep}{3pt}
\begin{tabular}{|l||c|c|c|}\hline
Schemes                              & Description                             & Power (Watt) \\\hline\hline
CryptoLight~\cite{Zheng:NANOARCH2022}& 8 photonic units, 21MB on-chip SPM      & 68.5        \\\hline
MATCHA~\cite{Jiang:DAC2022}          & ASIC, 8-core 4MB SPM, 8GB HBM2          & 37           \\\hline
FPT~\cite{Beirendonck:ARIX2022}      & Alveo U280, 8GB HBM, 16GB DDR4          & 99           \\\hline
Strix~\cite{Putra:MICRO2023}         & 8-core, 21MB on-chip SPM, HBMe          & 77.14           \\\hline
\end{tabular}
\label{t:t_hard_comp}
\end{table}

\begin{table}[ht!]
\vspace{-0.15in}
\centering
\caption{The parameters of DTFHE.}
\footnotesize
\setlength{\tabcolsep}{3pt}
\begin{tabular}{|l||c|c|c|c|}\hline
Parameters                   &  I          & II         & III       & IV      \\\hline\hline
$q$ of $\mathbb{Z}_q$        &  $2^{32}$   & $2^{32}$   & $2^{64}$  & $2^{128}$      \\\hline
Security Level ($\lambda$)   &  128-bit    & 110-bit    & 128-bit   & 128-bit      \\\hline
TLWE Dim. ($n$)              &  586        & 500        & 632       & 829       \\\hline
TGLWE Dim. ($k$)             &  2          & 1          & 1         & 1       \\\hline
Polynom. Size ($N$)          &  512        & 1024       & 2048      & 4096    \\\hline
Decomp. Base ($\beta$)       &  8          & 10         & 9         & 2       \\\hline
Decomp. Level ($l$)          &  2          & 2          & 4         & 1       \\\hline
\end{tabular}
\label{t:t_tfhe_para}
\vspace{-0.15in}
\end{table}

\textbf{Applications}. Besides DTFHE operations, we also study encrypted general-purpose applications implemented by TFHE and DTFHE. We selected four circuit designs from the ISCAS'85 benchmark suite. These circuits cannot be realized using traditional FHE schemes like CKKS, as they necessitate linear operations such as additions, as well as Boolean Algebra operations like XOR. Although TFHE can implement these circuits bit by bit, we demonstrate that DTFHE can efficiently and inherently support both linear and Boolean operations within the circuits.

\textbf{Simulation}. We simulated the OFHE's cycle-level performance by our in-house CGRA modeling framework, and validated it against Strix~\cite{Putra:MICRO2023}. Underpinning the architectural description of OFHE, the framework compiles DTFHE operations into data flow graphs, orchestrating and mapping each to an OFHE hardware unit. This methodology informs the derivation of both latency and energy metrics for each DTFHE operation.

\vspace{-0.1in}
\section{Results and Analysis}
\label{s:results}

\subsection{Comparison Against Prior Accelerators}

\begin{figure*}[t!]
\centering
\begin{minipage}[c]{0.2\textwidth}
\centering
\includegraphics[width=1.33in]{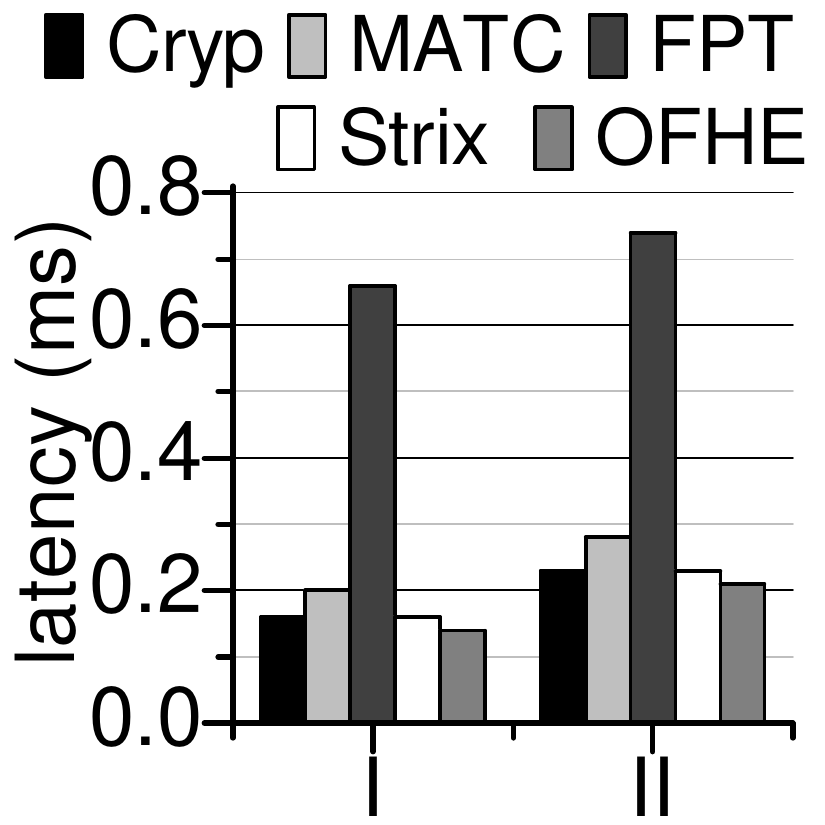}
\caption{GFB Lat.}
\label{f:ofhe_fb_lat}
 \end{minipage}
\hspace{-0.1in}
\begin{minipage}[c]{0.2\textwidth}
 \centering
\includegraphics[width=1.33in]{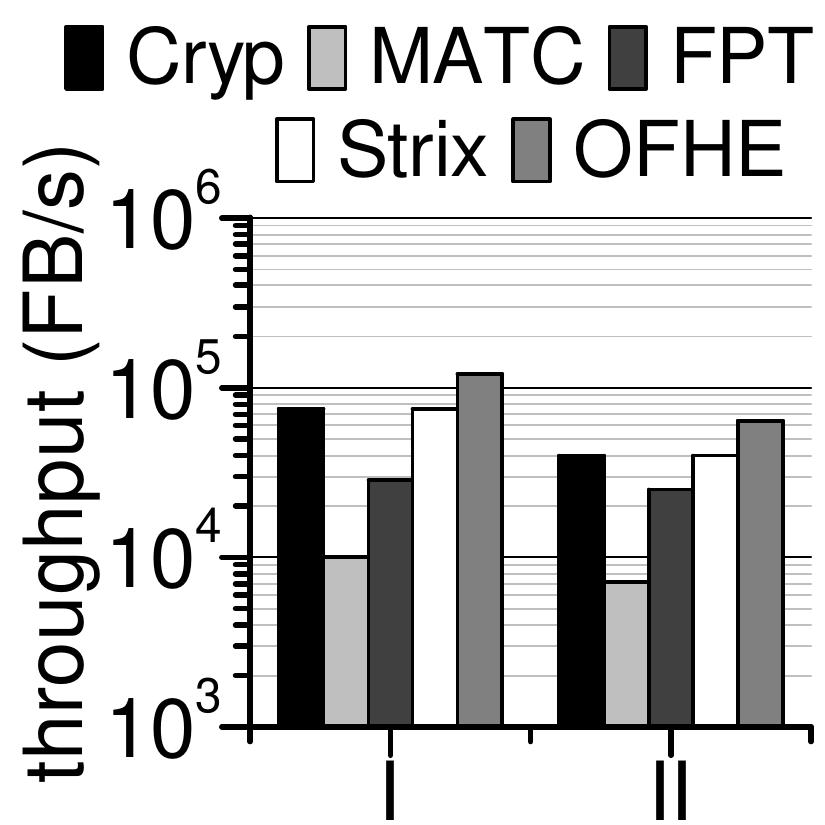}
\caption{GFB thr.}
\label{f:ofhe_fb_thr}
\end{minipage}
\hspace{-0.1in}
\begin{minipage}[c]{0.2\textwidth}
 \centering
\includegraphics[width=1.33in]{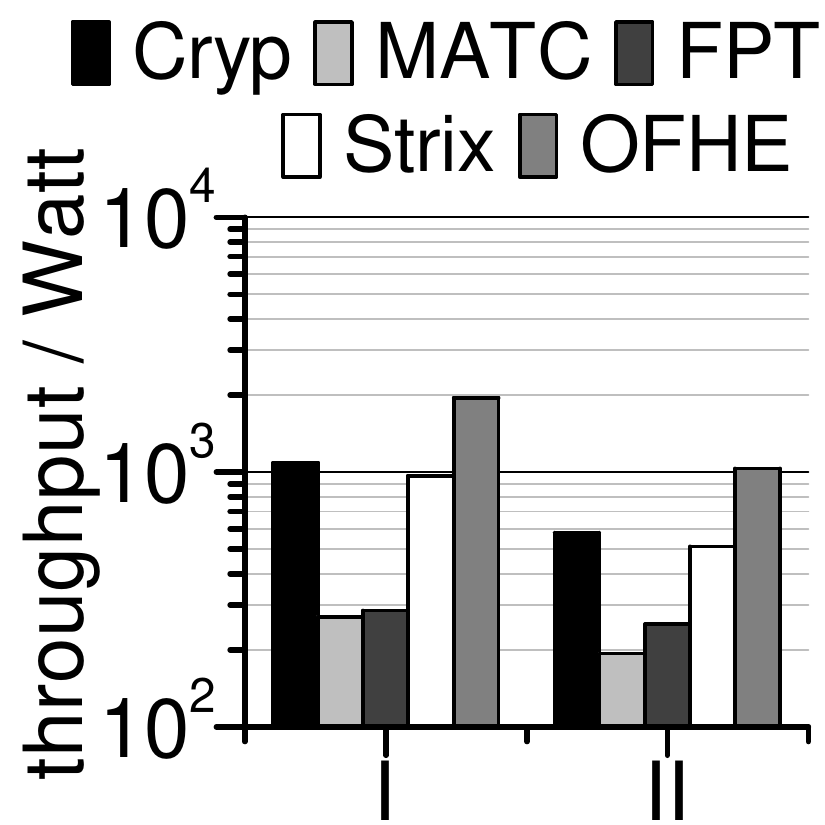}
\caption{GFB thr/W.}
\label{f:ofhe_fb_thw}
\end{minipage}
\hspace{-0.1in}
\begin{minipage}[c]{0.2\textwidth}
 \centering
\includegraphics[width=1.33in]{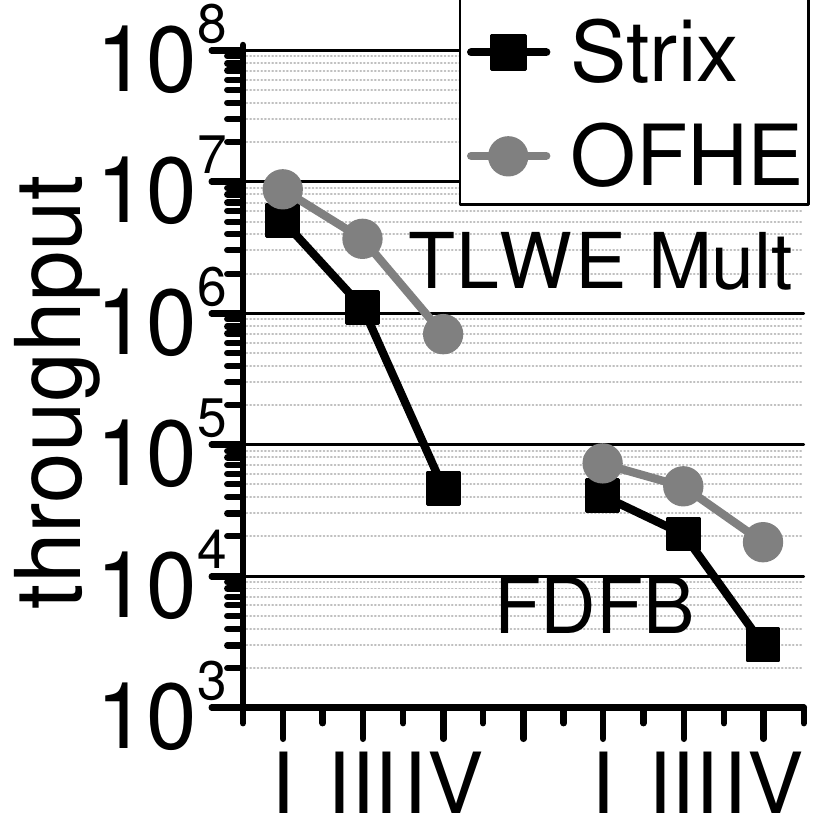}
\caption{DTFHE Ops.}
\label{f:ofhe_other_latency}
\end{minipage}
\hspace{-0.1in}
\begin{minipage}[c]{0.2\textwidth}
 \centering
\includegraphics[width=1.33in]{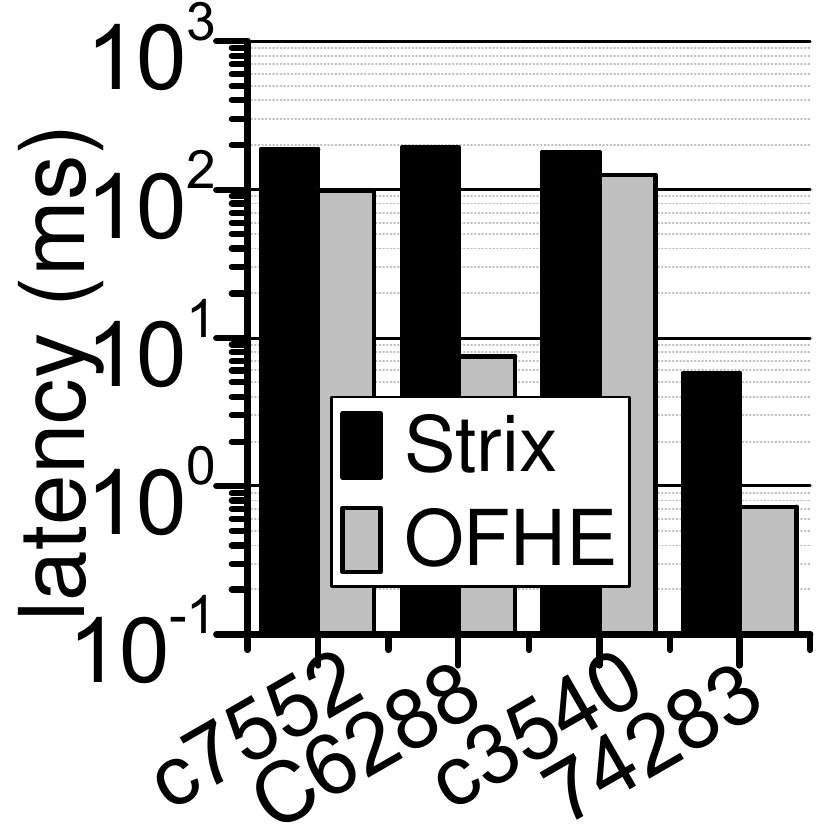}
\caption{DTFHE Apps.}
\label{f:ofhe_other_app}
\end{minipage}
\vspace{-0.15in}
\end{figure*}

\textbf{GFB Latency}. The bottleneck in TFHE gates with dual 1-bit inputs is the GFB, typically positioned at the end of a logic gate, operating on a 32-bit datapath. Figure~\ref{f:ofhe_fb_lat} illustrates the latency differences between prior accelerators and OFHE when executing a GFB. FPT exhibits the longest GFB latency for parameter sets I and II, while OFHE achieves the most significant latency reduction, trimming Strix GFB latency by 12.5\% and 8.7\% for parameter sets I and II, respectively. Although Cryp uses faster photonic adders and multipliers, its latency for a GFB is determined by its on-chip CMOS memory speed. During an (I)FFT kernel, Cryp needs to write every intermediate result into the on-chip memory, so it has almost the same latency as Strix.

\textbf{GFB Throughput}. As depicted in Figure~\ref{f:ofhe_fb_thr}, the GFB throughput varies significantly among accelerators. MATCHA and FPT face constraints in supporting numerous simultaneous GFB operations due to hardware limitations or lack of a fully-pipelined design. OFHE's PFFTE chips outperform FPT in GFB throughput, with enhancements of 61\% \& 57\% over Strix for parameter sets I and II.

\textbf{Throughput per Watt}. Figure~\ref{f:ofhe_fb_thw} depicts the energy efficiency of the accelerators, measured as GFB throughput per Watt. Apart from Cryp, the energy efficiency of the other platforms closely follows their throughput results. Cryp utilizes low-power photonic computing units, resulting in slightly superior energy efficiency compared to Strix. FPT, operating on FPGA substrate, incurs substantial power overhead, significantly reducing its energy efficiency. OFHE's PFFTE chips consume minimal power, resulting in a $2\times$ and $94\%$ improvement in GFB throughput per Watt compared to Strix for parameter sets I and II, respectively.

\vspace{-0.2in}
\subsection{Performance of 64- \& 128-bit Datapaths}

Figure~\ref{f:ofhe_other_latency} compares the throughput of GFBs, TLWE Mults, and FDFBs between OFHE and Strix. GFBs, TLWE Mults, and FDFBs utilize parameter sets I, III, and IV, respectively, to encrypt different sizes of plaintexts. For parameter sets III and IV, DTFHE requires 64- and 128-bit datapaths. Prior TFHE accelerators face a limitation as none support datapaths exceeding 32-bits. Consequently, Strix only natively supports parameter set I. We employ the Karatsuba algorithm~\cite{Guimaraes:CEA2022}, with a time complexity of $\mathcal{O}(3^{\log_2n})$ on Strix to process 64- and 128-bit operations using native 32-bit operations, where $n$ can be 64 or 128. This leads to an exponential decrease in throughput on Strix when transitioning from a 32-bit to a 64-bit to a 128-bit datapath, due to the high complexity of the Karatsuba algorithm. In contrast, OFHE maintains a linear latency evolution across its datapaths. Consequently, a significant throughput disparity arises between Strix and OFHE, ranging from $71.5\%$ to $35.6\times$.

\vspace{-0.2in}
\subsection{General Purpose Application Performance}
We selected four circuits from the ISCAS'85 benchmark: c7552 (32-bit adder/comparator), c6288 (16x16 multiplier), c3540 (8-bit ALU), and 74283 (4-bit adder). Figure~\ref{f:ofhe_other_app} compares the latency of these circuits on Strix and OFHE. While TFHE processes the circuits gate by gate on Strix, DTFHE utilizes multi-bit additions and multiplications on OFHE to realize the same circuits. OFHE demonstrates a latency reduction ranging from $29.4\%$ to $96.1\%$ compared to Strix. This improvement is particularly significant for c6288 and 74283, where most Boolean logic gates can be combined into multi-bit multiplication or addition operations. The superiority of DTFHE over TFHE in processing general-purpose applications is evident. For c7552 and c3540, the latency reduction benefits from both DTFHE advantages and OFHE's support for TLWE Mults and FDFBs. On average, compared to Strix, OFHE decreases the latency of these applications by 58.6\%.

\vspace{-0.1in}
\section{Conclusion}
\label{s:con}

This study introduces OFHE, an electro-optical accelerator designed for DTFHE operations on multi-bit plaintexts, particularly BFV-like TLWE multiplications and FDFBs. Leveraging PFFTE chips, OFHE efficiently addresses the FFT and IFFT kernel bottlenecks, achieving enhanced performance with reduced power. When benchmarked against existing TFHE accelerators, OFHE showcases a 8.7\% latency reduction, 57\% throughput improvement, and an $94\%$ boost in throughput per watt.

\begin{acks}
This work was supported in part by NSF CCF-2105972.
\end{acks}

\bibliographystyle{ACM-Reference-Format}
\bibliography{homo}

\end{document}